# Superconductivity in Ca-doped graphene

J. Chapman[1], Y.Su[1], C. A. Howard[2], D. Kundys[1], A. Grigorenko[1], F. Guinea[1], A. K. Geim[1], I.V. Grigorieva[1], R. R. Nair[1]

[1]School of Physics and Astronomy, University of Manchester, Manchester M13 9PL, UK

[2]Department of Physics and Astronomy, University College London, London, WC1E 6BT, UK

**Graphene, a zero-gap semimetal, can be transformed into a metallic, semiconducting or insulating state by either physical or chemical modification[1-3]. Superconductivity is conspicuously missing among these states despite considerable experimental efforts as well as many theoretical proposals[4-6]. Here, we report superconductivity in calcium-decorated graphene achieved by intercalation of graphene laminates that consist of well separated and electronically decoupled graphene crystals. In contrast to intercalated graphite[7], we find that Ca is the only dopant that induces superconductivity in graphene laminates above 1.8 K among intercalants used in our experiments such as potassium, caesium and lithium. Ca-decorated graphene becomes superconducting at $\approx$6 K and the transition temperature is found to be strongly dependent on the confinement of the Ca layer and the induced charge carrier concentration. In addition to the first evidence for superconducting graphene, our work shows a possibility of inducing and studying superconductivity in other 2D materials using their laminates.**

Nearly all allotropes of carbon including fullerenes, nanotubes, diamond and graphite were shown to exhibit superconductivity under heavy doping[7-10]. It is striking that no superconductivity has so far been observed in graphene, the basic building block for many of these allotropes. Interest in carbon-based superconductors has recently[11-13] been revived by the discovery of superconductivity in $CaC_6$, Ca-intercalated graphite compound (Ca-GIC) with $T_c \approx 11.5K$. Although some aspects of this superconductivity remain under debate[14-18], main contributing factors have been identified[14,15,17] as (i) doping via metal adatoms to reach sufficiently high electron concentrations in graphite, (ii) importance of an interlayer (IL) electronic band that comes from the intercalant superlattice formed between graphene layers, and (iii) the overall electron-phonon coupling that is related to coupling involving carbon phonons and intercalant vibrations.

According to recent DFT calculations[4], similar conditions are required to induce superconductivity in metal decorated graphene, i.e. doping adatoms are required not only to achieve sufficiently high electron concentrations, but also to create an electronic band arising from the adatom superlattice and ensure its overlap with the graphene π* band[4]. However, the effect of metal adatoms on free-standing graphene is predicted to be different from that in intercalated graphite. The difference is due to the quantum confinement of dopants' wave functions in the latter, and the absence of such confinement is expected to shift the IL band towards the Fermi level, thereby suppressing superconductivity in Ca-decorated graphene but enhancing it for Li doping[4]. Experimentally, previous attempts to induce superconductivity in graphene were limited to the proximity induced superconductivity[19] and *in situ* ARPES measurements on metal decorated graphene[20,21] which identified features attributed to dopant–related vibrational modes[20] and found signatures of heavy doping as well as the appearance of an IL band in Ca-intercalated graphene bilayer (no IL band could be seen for Li intercalation). Due to the nature of these experiments, however, they could not provide evidence of the emergence of intrinsic superconductivity.

In this report, we have investigated the possibility of inducing superconductivity in graphene by coating its crystallites with K, Cs, Li and Ca. To this end, we used so-called graphene laminates (GLs) that consist of graphene crystals arranged in a layered manner, similar to bulk graphite. However, unlike in graphite, crystallites within a GL are rotationally disordered and exhibit larger interlayer separations. This is known to result in effective decoupling of individual layers so that their electronic band structure corresponds to that of isolated graphene[22]. Accordingly, GLs offer a valuable alternative to individual graphene crystals in superconductivity studies because GLs can be produced in bulk and, therefore, measured using SQUID magnetometry, a method of choice for detecting superconductivity. In addition, bulk samples consisting of graphene and alkali monolayers are much less susceptible to environmental damage that arises due to extreme reactivity of alkali metals with oxygen, moisture, etc. We have employed different types of graphene laminates: those made directly from graphite (GLs), reduced graphene oxide laminates (RGOLs) and laminates containing both graphene and boron nitride (GBNLs). Samples were prepared using previously reported techniques[22-24] (Methods). To insert metal atoms between graphene crystallites within the laminates we employed techniques similar to those used previously for graphite intercalation (Methods). The effect of metal insertion was immediately obvious from visual inspection. Similar to intercalated graphite[7], GLs exhibited a pronounced colour change arising from changes in electronic structure upon doping (Fig.1a-d). As discussed below, different colours of metal-intercalated GLs correspond to different plasmon energies due to different doping levels.

Fig.1e shows typical magnetisation vs temperature curves, $M(T)$, for Ca-GL and Li-GL. Zero field cooling (ZFC) data for Ca-GL clearly shows a diamagnetic transition at ≈ 6.0 K (shielding of the external field, $H$, which is characteristic of superconducting materials). The large diamagnetic response corresponds to bulk superconductivity (superconducting fraction close to 100%), i.e., all constituent graphene crystallites become superconducting. The onset transition temperature found from $H=0$ $M(T)$ curves is $T_c^{onset}(0) = 6.4\pm0.4$ K, varying only slightly from sample to sample. The relatively broad superconducting transition is likely to be due to either different levels of doping for individual graphene crystallites or disorder in Ca monolayers (see below). The inset in Fig. 1e shows evolution of $M(T)$ with increasing $H$: both the diamagnetic response and $T_c$ decrease as expected. The $T$-dependent upper critical field, $H_{c2}(T)$, shown in Fig. 1f exhibits a positive curvature consistent with temperature-dependent critical fields for superconductors made of weakly coupled superconducting layers, such as alkali-metal intercalated $MoS_2$ and $Bi_2Sr_2CaCu_2O_8$[25] and no detectable anisotropy (Supplementary information). In contrast, neither K-, Cs- or Li-intercalated GLs showed any sign of a superconducting transition down to our lowest temperature of 1.8 K (red symbols in Fig.1e and Supplementary information). Therefore, below we focus on Ca-decorated GLs only.

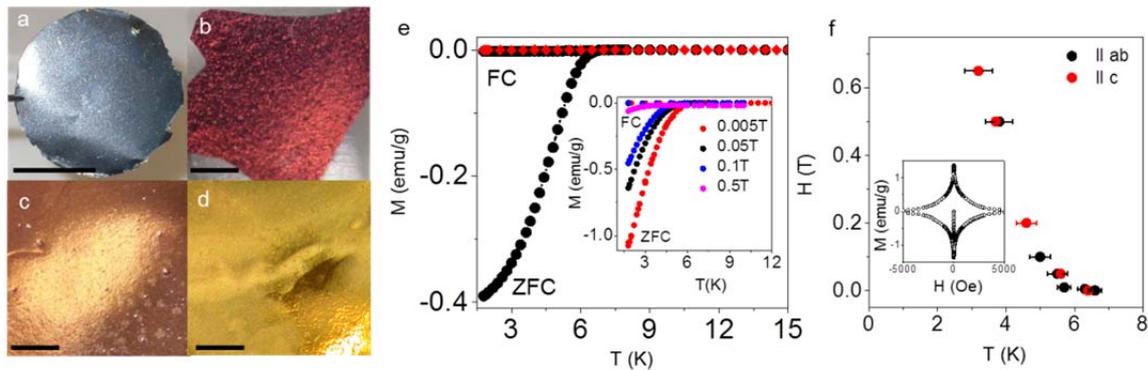

*Figure 1. Characterisation of intercalated graphene laminates. (a-d)* Optical photographs of pristine (Scale bar 1 cm), K-, Cs- and Ca- intercalated GLs (Scale bar 1 mm), respectively. Li-GL (not shown) has a similar colour to Ca-GL. (*e*) Temperature dependence of ZFC and FC mass magnetisation, M, for Li-GL (red symbols) and Ca-GL (black) at H = 4 Oe applied parallel to the laminates' surface. The estimated systematic error in determining M is ~10% due to inaccuracy of measuring the sample mass that was typically several mg (this is difficult because of extreme sensitivity of intercalated GLs to moisture and oxygen). The inset shows ZFC and FC M(T) at different H for Ca-GL. (*f*) Main panel: phase diagram for Ca-GL obtained from the M(T) plots for different H applied parallel (black dots) and perpendicular (red) to the graphene plane. The inset shows the magnetisation dependence as a function of H // ab, which is characteristic of type-II superconductors with significant trapping of magnetic flux (pinning).

Further evidence for superconductivity in Ca-GLs was obtained from behaviour of their electrical resistivity, $R(T)$ (see Fig. 2a). The zero-field resistive transition is rather broad with an onset at $T_c \approx$ 12K. The much higher $T_c$ compared to that found in $M(T)$ measurements indicates sample inhomogeneity, possibly due to the presence of some intercalated few-layer graphene (effectively ultrathin intercalated graphite; its $T_c$ is expected to be similar to bulk graphite, $T_c^{bulk} \approx$ 11.5K[11,12]). We emphasise that the fraction of few-layer graphene in our GLs is very small, and this higher-$T_c$ phase could not be discerned in our $M(T)$ measurements.

To find out whether the observed superconducting response corresponds to a bulk layered system similar to intercalated graphite or, alternatively, is representative of superconductivity within individual graphene crystals, we have prepared mixed laminates where graphene crystallites are interspersed with BN flakes (GBNLs), and RGOLs where graphene flakes have larger separations compared to GLs, ≈ 3.6 Å versus ≈ 3.4 Å (Supplementary information). The mixed laminates were then intercalated with Ca using the same method as above. By adding extra BN layers in GBNLs, graphene crystallites were physically separated from each other. For example, in a 1:1 (weight) mixture of graphene and BN, statistically most of graphene crystals should have BN rather than graphene as its nearest neighbour. For higher concentrations of BN, graphene flakes are separated even further.

Ca intercalation of GBNLs and RGOLs was again evident from colour changes: In contrast to golden Ca-GL, Ca-RGOL is metallic brown whereas Ca-GBNLs' colours varied from metallic brown to metallic green/blue with increasing BN content (Fig.3a-d). We have found that Ca-RGOLs and Ca-GBNLs exhibit superconducting characteristics practically identical to those of Ca-GL, and the only pronounced difference is a reduction in $T_c$ (Fig 2b,c). Specifically, $T_c$ for Ca-RGOL is reduced by ≈ 2 K and, for Ca-GBNLs, it decreases monotonically from ≈ 6.4 K to ≈ 4.4 K with increasing BN content up to 70%. Importantly, the addition of BN did not change either the width of the superconducting transition, or the superconducting fraction normalised to the graphene content (Fig. 2c). This strongly indicates that the superconductivity arises from independent Ca-decorated graphene crystals.

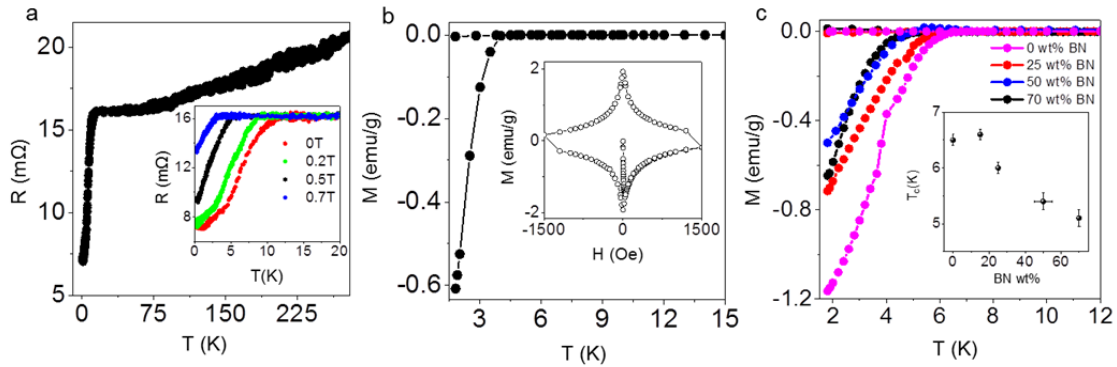

*Figure 2. Superconductivity in Ca-GL, Ca-RGOL and Ca-GBNLs. (a) Temperature dependence of the electrical resistivity of a 3 µm thick 3×3 mm sample of Ca-GL showing a superconducting transition at ≈12K. The inset shows the evolution of R(T) with increasing external magnetic field, H. The sample did not reach zero-resistance state, probably as a result of partial degradation because of brief exposure to air during transfer into a cryostat (Methods). (b) Temperature dependence of ZFC and FC magnetisation for Ca-RGOL at 4 Oe applied parallel to the graphene plane. The inset shows an example of the corresponding M(H); T=1.8 K. (c) Magnetisation of Ca-GBNLs with different BN contents (M is normalised to the graphene content, i.e. only the mass of graphene is included for each GBNL). The inset shows the dependence of $T_c$ on BN concentration (weight %).*

To understand the origin of different $T_c$'s in Ca-GL, Ca-RGOL, and Ca-GBNLs and relate these to their electronic structures, we used X-ray analysis, Raman spectroscopy and optical reflectivity measurements to probe the laminates' structure, phonons and plasmons, respectively. X-ray analysis revealed that the average separation of graphene layers in Ca-GLs and Ca-RGOLs is significantly larger than the interlayer spacing in the corresponding graphite intercalation compound (Ca-GIC): $d$ ≈ 5.1 and 5.4 Å vs 4.5 Å (ref. [12]), respectively, presumably due to weaker coupling between individual corrugated graphene crystallites within GLs. According to theory[4], the IL band that forms as a result of metal deposition is sensitive to the separation between graphene layers, which changes dopants' wavefunctions because of the quantum confinement. Furthermore, the increased $d$ in Ca-GLs effectively reduces the overlap between the $\pi^*$-band of graphene and the IL band, which reduces both charge carrier concentration and electron-phonon interactions, thereby reducing $T_c$[4,26].

To estimate charge carrier concentrations, $n$, in Ca-GL, Ca-RGOL, Ca-GBNLs and, compare them with that in Ca-GIC, we measured optical reflectivities of these compounds (Fig. 3e). The clear shift of the reflectivity minima to lower energies indicates a reduction in plasmon energy, $\omega_p$ (Supplementary information) or – equivalently – a reduction in the overall electron concentration, $n$. The plasmon energy is also related to the observed changes in visual colour of the compounds (cf. Figs. 1 and 3). One can see that the reduction in $n$ is accompanied by progressively lower $T_c$ (Supplementary information).

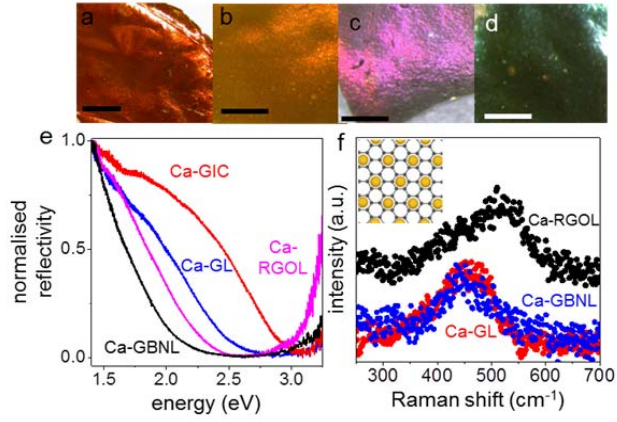

*Figure 3. Optical characterisation of different Ca-doped laminates. (a-d) Photographs of Ca-RGOL and Ca-GBNLs with 25, 50 and 70 wt% BN content, respectively. Scale bars, 1 mm. (e) Reflectivity spectra for Ca-intercalated graphite and graphene laminates. (f) Out-of-plane Raman mode for different laminates. Ca-GBNL (blue) had 50 wt% of BN. The inset shows a sketch of the Ca superlattice on graphene; large yellow circles are Ca atoms and small black dots are carbon atoms.*

Plasmon energies were determined by fitting the reflectivity curves with the expression for the reflection coefficient for metallic systems (Supplementary information). To extract 2D carrier concentrations from the measured $\omega_p$, we used a model where the metal-graphene layers are represented by electrostatically coupled two-dimensional units (Supplementary information). A similar model was successfully used in ref. [26] to explain the empirical correlation between the filling of the IL band and the occurrence of superconductivity in GICs. This yielded the following relation between $\omega_p$, electron concentrations in graphene and the IL band, $n_c$ and $n_{IL}$, and the layer separation $d$:

$$\hbar^2\omega_p^2 = \frac{2e^2}{\epsilon_0 d}\left(\frac{2\pi\hbar^2 n_{IL}}{m_{IL}} + 2\sqrt{3\pi}\hbar v_F\sqrt{|n_C|}\right),$$

where $e$ is the electron charge, $\epsilon_0$ the permittivity of free space, $v_F$ the Fermi velocity in graphene and $m_{IL}$ the mass of the metal ions. Unlike in a bulk metal where $\omega_p$ is determined only by the total carrier density, plasmon energies in layered systems also depend on the distribution of electrons between graphene and the IL band, and on $d$. For example, for Ca-GIC, Ca-GL, and Ca-RGOL, respectively, we obtain $n \approx$ 1.8×10$^{14}$, 1.1×10$^{14}$ and 9×10$^{13}$ cm$^{-2}$ (Supplementary information). According to the Bardeen-Cooper-Schrieffer (BCS) theory, such changes in $n$ alone could in principle account for the observed differences in $T_c$. For example, a ~10% reduction in $n$ between Ca-GL and Ca-RGOL should reduce $T_c$ by ~2K (Supplementary information). This is in agreement with our observations (e.g., 6.4K for Ca-GL and 4K for Ca-RGOL). However, $\omega_p$ and $n$ for Ca-GBNLs are lower than for Ca-RGOLs, in contrast to the opposite relation between their $T_c$. Similar comparison between Ca-GIC and Ca-GL (~30% reduction in $n$, see supplementary information) suggests much larger suppression of $T_c$ than observed. All this indicates that $n$ is not the only factor at play. Furthermore, Li- and Ca- GLs had equal plasmon energies (Supplementary Table 1) and similar $n$ but no superconductivity could be detected for Li-GL, similar to Li-GIC that is not superconducting[7]. The reason for so different superconducting properties of equally doped compounds has been suggested before[14,15,18] as either occupied or unoccupied IL bands in Ca-GIC and Li-GIC, respectively. Our experiment highlights the fact that the same doping can result in different distributions of charge carriers between the graphene (Dirac) and IL bands. We believe that in the case of Ca-GL the IL band is occupied but for Li-GL it remains empty, similar to the case of Li-GIC.

Further information about relative contributions of Dirac and IL bands comes from Raman spectroscopy. For bulk Ca-GIC, Raman spectra are known[27,28] to have two main features: an in-plane bond-stretching mode at ~ 1500 cm$^{-1}$ and a weaker ~ 450 cm$^{-1}$ mode due to out-of-plane vibrations. The latter originates from folding of the K-point graphene phonon to the $\Gamma$ point in the larger unit cell defined by the √3 x √3 Ca superlattice. The 450 cm$^{-1}$ mode has been shown to be sensitive to separation between graphene layers in different GICs[28]. As the layer separation increases, this mode red-shifts, concomitant with the observed decrease in $T_c$. We have found the out-of-plane mode for all the Ca-intercalated GLs (Fig. 3f) which confirms the presence of the √3 x √3 Ca superlattice (inset in Fig. 3f). The relatively broad peaks in Fig. 3f compared to Ca-GIC[27] indicate notable disorder in the Ca superlattice, possibly at graphene edges. In contrast to Ca-GIC where the out-of-plane mode is at ≈ 440 cm$^{-1}$, the corresponding Raman peaks for Ca-GL and Ca-RGOL are red-shifted to ≈ 460 and 520 cm$^{-1}$, respectively. A red shift of this phonon mode compared to bulk GIC has been predicted[4] for Ca-decorated monolayer graphene to occur due to a weaker confinement of the Ca layer. Our Raman data indicate the progressively weaker confinement from Ca-GIC to Ca-GL to Ca-RGOL, consistent with their increasingly larger interlayer distances 4.5 Å to 5.1 Å to 5.4 Å, respectively. The position of the out-of-plane mode for Ca-GBNLs is similar to that of Ca-GLs (Fig.3f) indicating comparable Ca confinement. Accordingly, the differences in $T_c$ in these cases can be attributed to decreasing $n$ with increasing a BN content, as evident from the reflectivity measurements discussed above as well as from the corresponding red shifts of the in-plane Raman mode (Supplementary information).

In conclusion, we have shown that graphene crystals decorated with Ca exhibit robust superconductivity with a transition temperature governed by the electron transfer from the metal to graphene and by the Ca-layer confinement that dictates the overlap between the IL and graphene electronic bands. In contrast to theoretical predictions, no superconductivity could be detected above 1.8K for the case of Li doping, possibly because the Li layer confinement in our materials was still too strong. Curiously, as plasmon energies of Ca-doped graphene lie in the visible range, sample colours can be used as a simple guide to estimate $T_c$.

Methods

Pristine graphene laminates (GLs) were fabricated as reported earlier[22,23,29]. In brief, high purity HOPG crystals were exfoliated in N-Methyl-2-pyrrolidone (NMP) in an ultrasonic bath and the resulting dispersions centrifuged at 12,000 rpm to obtain a stable suspension. These were then filtered through porous alumina filters to obtain several μm thick free standing laminates of graphene. Reduced graphene oxide laminates were also prepared as reported previously[24] (details in Supplementary information). Recent progress in reducing graphene oxide back to graphene[24,30] allows synthesis of high quality RGO with few defects. In the present work GO was reduced using hydroiodic acid. GBNLs were prepared by the same method as GLs but with the filtration of composite solution of graphene-BN suspension in NMP (Supplementary information).

Metal intercalation was done in either high vacuum or an argon-filled glove box to avoid exposure of the highly reactive alkali-/alkali-earth metals and the intercalated samples to ambient moisture and oxygen. We have used both pure-metal vapour transport[7,11] and alloy-intercalation techniques[12] to insert K, Cs, Li and Ca into GLs (Supplementary information).

Magnetisation measurements were performed on 4 × 4 mm square samples using Quantum Design MPMS XL7 SQUID magnetometer (Supplementary information). In the zero field cooling (ZFC) mode, the samples were initially cooled to 1.8 K in zero applied field, then a desired external field $H$ applied and the magnetisation $M$ measured as a function of increasing temperature, $T$ (typically 1.8 - 30 K). The field-cooling (FC) part of an $M(T)$ curve was obtained on cooling the sample to 1.8 K in the same $H$.

For electrical transport measurements, we have fabricated GL devices in van-der-Pauw geometry, i.e. four contacts were made with silver paint in the corners of a 3 × 3 mm square sample of a graphene laminate. The devices were then intercalated with Ca using vapour transport technique, transferred to a container inside the glove box and quickly cooled down to liquid nitrogen temperature to avoid degradation of the sample. Later, the samples were transferred to a liquid helium cryostat and cooled down to 0.3 K and the resistance of the device continuously monitored while cooling. All transport measurements were performed using standard four probe DC measurement techniques using Keithley's 2400 source-meter and 2182A nanovoltmeter.

Raman spectra were acquired using a Renishaw micro Raman spectrometer with a 514 nm excitation using a laser power <1 mW. Due to the extreme sensitivity of the samples to air, they were sealed inside quartz tubes in the inert atmosphere of a glove box to avoid degradation.

Optical reflectivity measurements were carried out using an Energetiq laser-driven light source (available wavelength range 190nm-2.4μm), where the light passed through a broadband fibre into a reflective collimator, a neutral density filter wheel and a 70/30 beam splitter before being focused with a 25x objective (NA – 0.65) onto the sample. To prevent degradation of the samples, they were covered with a thin film of paraffin oil and sealed inside a glass cell in the inert atmosphere of a glove box. To eliminate reflection from the glass plate encapsulating the sample, we used a refractive index matching gel. The sample was brought into focus with a three-dimensional stage manipulation system. The reflected light passed back through the lens and beam splitter, before being split again (92/8) to allow a digital image to be simultaneously captured by the camera (8%), with the remaining light (92%) then focused along another broadband fibre to the Ocean Optics spectrometer for analysis. The obtained digital images allowed us to confirm that the samples remained stable (did not degrade) during the measurements by monitoring their colour. The spectra were taken at 345-1040 nm wavelengths, using a silver mirror as a reference.

# Supplementary Information

# Superconductivity in Ca-doped graphene

**Supplementary Note 1. Preparation of reduced graphene oxide laminates (RGOLs)**

To produce graphite oxide we used high-purity crystals of highly-oriented pyrolytic graphite (HOPG). The crystals were broken into small pieces and oxidised using a modified Hummers' method[1], with all oxidation reactions carried out below 10 $^{\circ}$C to minimise formation of defects during the reactions. This method has been shown to produce – after chemical reduction - high quality graphene with lower amounts of defects than conventional Hummer's method[2], as demonstrated by high carrier mobilities in corresponding devices (>1000 cm$^2$V$^{-1}$s$^{-1}$)[1]. Graphite oxide flakes were exfoliated in water in an ultrasonic bath and then centrifuged at 10000 rpm to separate monolayer graphene oxide (GO) flakes. The high degree of hydrophilicity of GO ensures very efficient exfoliation with nearly 100% yield of graphene oxide monolayers[3]. By adjusting sonication parameters we were able to controllably vary the typical size of individual GO crystallites between ~ 0.2 and 20 μm, but did not notice any effect of the crystallite size on the superconducting properties (after Ca intercalation). GO laminates (GOLs) were prepared from GO dispersions by filtration through alumina membranes with 20 nm pore size. As an additional measure to ensure the absence of metallic impurities in the samples prior to intercalation[4], all GOLs were immersed in concentrated HNO$_3$ for 24h before and after the chemical reduction.

To convert graphene oxide laminates into graphene laminates (RGOLs) we used chemical reduction[5,6] with hydroiodic (HI) acid. This method has been shown to produce higher quality RGO films and laminates, with fewer defects, than other reduction methods[7]. Reduction was carried out by immersing GO laminates in HI acid for 30 minutes, followed by repeated rinsing with ethanol to remove residual HI.

**Supplementary Note 2. Intercalation of graphene-based laminates and bulk graphite with K, Cs, Li and Ca**

To decorate graphene crystallites in the laminates with K, Cs and Li, we used a technique similar to the well-established vapour transport method[8]. To this end, a sample of GL and the chosen metal were placed inside a tantalum foil envelope and transferred to a stainless steel or glass tube sealed with a valve, all of this done in the high-purity argon atmosphere inside a glove box. The tube was then evacuated to ≈ 10$^{-6}$ mbar and the whole assembly heated in a furnace to an appropriate temperature (200°C for K and Cs and 350°C for Li) in order to vaporise the metal. The high vacuum in the reaction container was maintained by continuous pumping to prevent oxidation of the reactants or the products. After approximately 40 hours of heating / exposure to the metal vapour, the products were recovered in a dry argon atmosphere. In the case of Li intercalation of GLs, we found it necessary to use a lower temperature compared to intercalation of bulk graphite reported in literature[9] (350 $^{\circ}$C vs 400 $^{\circ}$C): at 400 $^{\circ}$C graphene partially reacted with Li to form lithium carbide that could be detected as an additional peak in the corresponding X-ray spectra. No carbides were formed in Li-GLs at 350 $^{\circ}$C. The reference bulk graphite samples (Li-GIC) were intercalated at 400 °C and did not show any lithium carbide signals in the X-ray data, in agreement with literature. For K and Cs intercalation, we used the same temperature (200 °C) for both GLs and the reference bulk graphite.

For Ca intercalation, we used both vapour transport[10,11] and alloy techniques[12]. In the former case, a GL sample was placed alongside an ingot of calcium metal inside a quartz tube evacuated to <$10^{-7}$ mbar and degassed at 350 °C for 24 hours in a set up similar to that in ref. [11]. The temperature was then increased to 470 °C in order to vaporise Ca metal. After exposing the GL sample to Ca vapour for one to two weeks, the Ca-GL sample was recovered from the container and stored in a dry argon atmosphere inside a glove box. For comparison, we have also prepared Ca-decorated mechanically exfoliated monolayer graphene and reference samples of Ca intercalated bulk graphite (Ca-GIC), using the same vapour transport method. To this end, a ~40 x 20 μm graphene monolayer was exfoliated onto an oxidised Si substrate by micromechanical cleavage and identified by optical contrast and Raman spectroscopy[13,14]. It was then exposed to Ca vapour with different exposure times (to vary the Ca coverage) and characterised using Raman spectroscopy (see below). Successful Ca coating of the graphene monolayer was also noticeable from an increase in optical contrast, similar to the colour changes observed for Ca-GLs.

An alternative intercalation method, the so-called alloy technique used previously for intercalation of Ca into bulk graphite[12], was employed to insert Ca into some of the GLs and all RGOLs and graphene-BN mixed laminates (GBNLs). To this end, GLs, RGOLs and GBNLs were exposed to molten calcium-lithium (≈ 20 at.% Ca) alloy at ~350 °C under dry argon atmosphere of a glove box for 12-18 hours. In addition to intercalation of Ca between the graphene crystallites in the laminate, this left a thin layer of metal on the sample surface, which was subsequently removed by gentle scratching with a ceramic scalpel. The superconducting properties ($T_c$, evolution of magnetisation with field and temperature) as well as Raman signatures of Ca-GLs prepared using the two different techniques (vapour transport and alloy intercalation) were identical, in agreement with earlier experiments on Ca-GIC[10,12].

**Supplementary Note 3. Magnetisation measurements**

Magnetic response of metal-intercalated samples was measured using a commercial SQUID magnetometer Quantum Design MPMS XL7. To prevent degradation of the samples during transfer to the cryostat and subsequent measurements, all samples were immersed in paraffin oil and sealed inside polycarbonate capsules in dry argon atmosphere of a glove box, then quickly transferred to the cryostat and immediately cooled down to below ~30K. The superconducting response of Ca-intercalated laminates is discussed in detail in the main text. In contrast to Ca intercalation, no superconductivity could be detected in Li-, K- and Cs intercalated GLs down to 1.8K (the lowest available temperature). Supplementary Fig. S1a shows ZFC and FC $M(T)$ for Li-GLs. The weak paramagnetic response obvious in Fig. S1a is not discernible on the scale of Fig. 1e in the main text where it is also shown as red symbols. Similar purely paramagnetic behaviour was observed for Cs-GLs (not shown). The high level of electron doping in Li- and Cs-GLs was evident from the emergence of Pauli paramagnetism – linear, temperature-independent $M(H)$ (Supplementary Fig. S1b) but it did not result in the emergence of superconductivity. For K-GLs, the magnetic behaviour was more complex (inset in Fig. S1a) showing hysteresis between ZFC and FC *M(T)*, possibly related to coupling between weakly magnetic K clusters, as suggested in ref. [15].

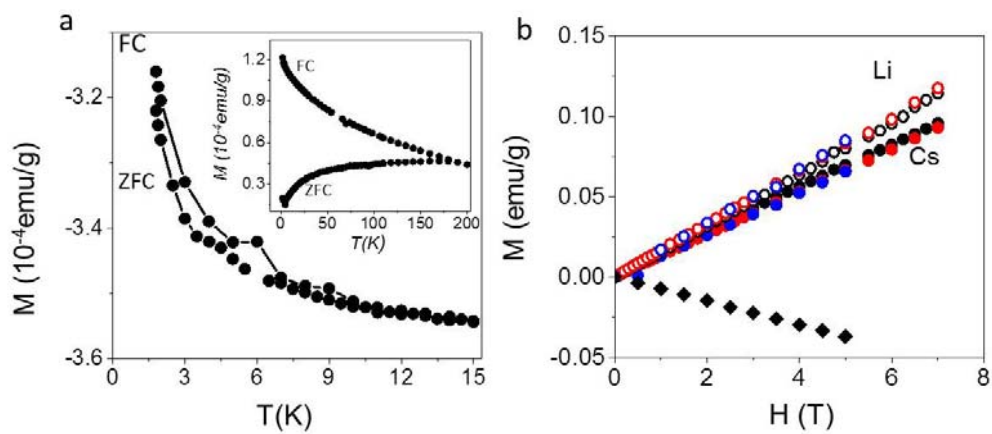

***Supplementary Figure S1.*** *(a) Main panel: ZFC and FC temperature-dependent magnetisation of Li-GL; H = 50 Oe. Inset: same for K-GL at H= 10 Oe. (b) Magnetic-field dependent magnetisation of Cs-GL and Li-GL showing a linear, temperature-independent response: black symbols: T=1.8 K; red symbols: T=10 K and blue symbols: T=100 K. As a reference, black diamonds show the diamagnetic response of GL before intercalation.*

The upper critical field, $H_{c2}(T)$, of Ca-GLs was extracted from measurements of $H$-dependent critical temperature, $T_c^{onset}$; the corresponding phase diagram is shown in Fig. 1f in the main text. Although the temperature interval, where the measurements could be made, is too narrow to attempt fits to theory and try to estimate $H_{c2}(0)$, the absence of anisotropy and the positive curvature due to weakly coupled superconducting layers[16] is clear from our measurements. The total absence of anisotropy (Fig. 1f) is surprising, as it is in contrast to the finite anisotropy of the bulk Ca-intercalated graphite, where the ratio, $\gamma$, of $H_{c2}$ parallel and perpendicular to the layers was founds to be $\gamma \approx 4$[17].

**Supplementary Note 4. GBN mixed laminates**

Crystals of h-BN purchased from Manchester Nanomaterials Ltd[18] were exfoliated in N-Methyl-2-pyrrolidone (NMP) using ultra-sonication as reported previously[19]. The dispersions were centrifuged at 12,000 rpm three times in order to remove multilayer flakes and to obtain a stable suspension of BN. After that composite graphene-BN suspensions were prepared by mixing graphene and BN suspensions in a desired proportion, followed by further ultra-sonication. A similar technique was used recently to prepare artificial van der Waals solids with electrical, mechanical, and optical properties distinctly different from those of the 'parent' layers[20]. Graphene-BN laminate samples (GBNLs) were then prepared in the same way as GLs, i.e. by filtration through an alumina membrane. The finished laminates were characterised using X-ray diffraction and Raman spectroscopy (see below). X-ray diffraction patterns for GBNLs with different BN content were similar to the pristine GLs, yielding the same layer spacing, $d \approx 3.4$Å. Such similarity is to be expected due to the nearly identical crystal lattices of h-BN and graphite. Intercalation of GBNLs with Ca was done using the alloy method as described above.

**Supplementary Note 5. X-ray diffraction**

To determine the interlayer separation in different graphene-based laminates we used X-ray diffraction. Similar measurements were used previously to measure the interlayer separation in Ca-GIC, which was found to be ≈ 4.5 Å[10,12]. Due to the sensitivity of Ca-GLs and Ca-RGOLs samples to air, they were sealed inside an airtight specimen holder transparent to X-rays (purchased from Bruker). Even with this protection, the environmental stability of the Ca-laminates was poorer than for intercalated bulk graphite, e.g., after multiple repeated scans the new peaks due to Ca insertion gradually disappeared, concomitant with a re-appearance of the peaks characteristic for pristine GLs. Therefore, all measurements were done as quickly as possible.

Supplementary Figure S2a,b highlights the shift in X-ray diffraction peaks corresponding to the interlayer separation in GL and RGOL before and after Ca intercalation. Before Ca insertion the interlayer separations in GLs and RGOLs were, respectively, $d ≈ 3.3\text{-}3.5$ Å and $3.6\text{-}3.8$ Å (Fig. S2a). The larger interlayer separation in RGOLs compared to GLs could be due to the GO reduction mechanism: During chemical reduction, oxygen-containing functional groups are removed as water or gas molecules and the release of these molecules from the interior of the laminate can induce corrugation and larger interlayer spacing compared to GLs[6]. The same diffraction peaks after Ca insertion is shown in Fig. S2b: the new peak positions correspond to $d ≈ 5.1$ Å for Ca-GL and $≈ 5.4$ Å for Ca-RGOL, i.e. the interlayer spacing in Ca-RGOL is still ≈0.3 Å larger that in Ca-GL, a similar difference as for the corresponding pristine samples. We note that the above diffraction peaks become markedly narrower after Ca intercalation (cf. panels (a) and (b) in Fig. S2). The reason for this is not clearly understood and difficult to investigate due to gradual sample degradation but may be an indication that the formation of Ca superlattice between corrugated and misaligned graphene flakes makes the layer structure more ordered in the direction perpendicular to the layers.

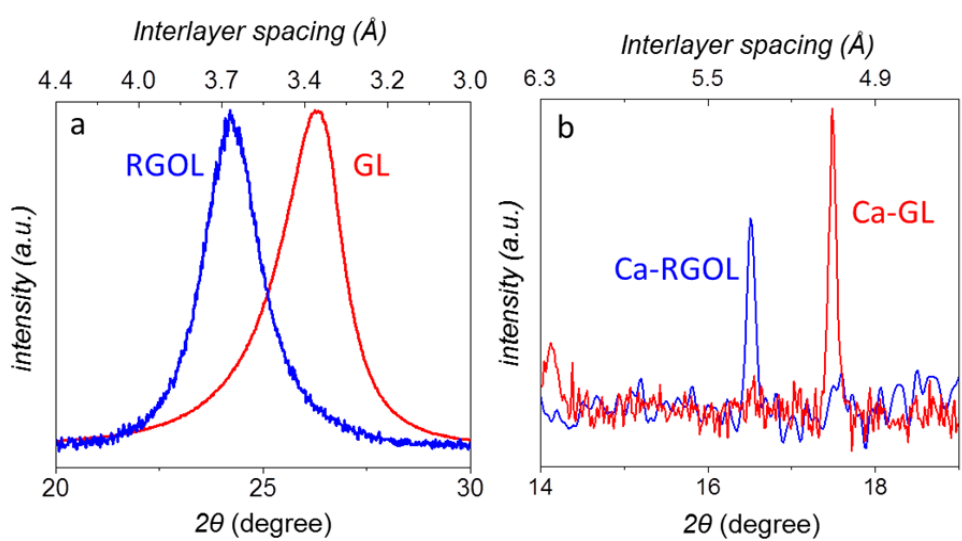

***Supplementary Figure S2.*** *X-ray diffraction peaks corresponding to the layer separation (a) in pristine GL and RGOL and (b) in Ca-GL and Ca-RGOL (background from the sample holder subtracted).*

**Supplementary Note 6. Raman spectroscopy**

Raman spectroscopy has been shown to permit qualitative understanding of the phonons and the degree of doping of graphene layers in intercalated graphite[11,21]. In particular, the level of doping of mono- and few-layer graphene exposed to K metal vapour was found to be continuously tunable due to increasing coverage with K atoms, until saturation is reached after a number of repeated exposures[11]. Such tunability was in contrast to bulk GICs that form distinct stoichiometric compounds[8].

In our work, we used Raman spectroscopy to probe the out-of-plane graphene phonons, as well as the level of doping, and to investigate the differences between Ca-GIC, Ca-GLs and the Ca-decorated monolayer graphene (MLG) exfoliated onto an oxidised silicon substrate (see Supplementary Note 2). Supplementary Figure S3a shows the evolution of Raman spectra for the MLG as Ca coverage is increased through repeated exposures to Ca vapour. For pristine graphene, the spectrum shows the expected single-component 2D band at ~ 2690 $cm^{-1}$ and the G-peak at ~ 1580 $cm^{-1}$. After intermediate Ca exposures, the 2D peak first decreases in intensity and then disappears altogether, due to raising of the Fermi level and the removal of the resonance conditions. This is accompanied by significant broadening and a blue shift of the G-peak that takes on a Breit-Wigner-Fano (BWF) line shape. Similar softening and broadening of the Raman modes have been observed previously on monolayer graphene decorated with K and were interpreted as evidence of high doping and increased electron-phonon interactions[11]. Some new features also appear between 1200 $cm^{-1}$ and 1400 $cm^{-1}$, similar to those found in K decorated monolayer graphene[11].

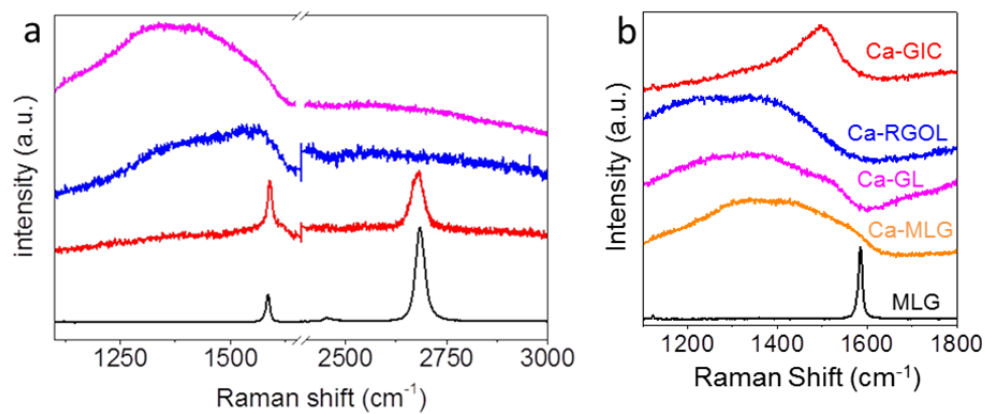

**Supplementary Figure S3.** *(a) Raman spectra of a monolayer graphene (MLG) exfoliated onto an oxidised Si substrate and repeatedly exposed to Ca vapour. Black curve: pristine state; red and blue curves: intermediate exposures; magenta curve: saturated Ca coverage. (b) Comparison of the Raman spectra for different Ca-decorated samples.*

The shape and position of the Raman G-peak for Ca-GL, Ca-RGOL and Ca-MLG are compared in Supplementary Fig. S3b. For comparison, we also show the Raman spectra of the pristine graphene (before exposure to Ca), and of the bulk Ca-GIC. In agreement with earlier measurements[22,23], the Raman spectrum for our Ca-GIC has a single BWF-shaped peak at ~ 1500 $cm^{-1}$, due to the Raman active $E_g$ mode corresponding to the in-plane bond stretching vibrations in graphene layers. The larger width and the blue shift of this peak with respect to pristine graphite (where G peak is found at ~1580 cm$^{-1}$) are believed to be due to an increase in electron-phonon interaction at large doping[11,23].

As clear from Fig. S3b, the Raman spectra of Ca-intercalated GL and Ca-RGOL are significantly more similar to the spectrum of the Ca-saturated MLG than to bulk Ca-GIC, indicating similar electronic properties of these three systems. These spectra serve as another indication that Ca-decorated graphene crystallites in Ca-GLs and Ca-RGOLs are effectively independent from each other and exhibit the same characteristics as similarly doped monolayer graphene.

**Supplementary note 7. Gradual doping and colour change.**

To investigate whether the doping level of graphene-based laminates can be tuned continuously, as in the case of the monolayer, or the exposure to Ca vapour/molten Li-Ca alloy results in the formation of a stoichiometric $CaC_6$ as in bulk graphite[10,12], we varied the time of exposure of GLs to Ca, Li, K and Cs metals. (We note that only the fully saturated first stage $CaC_6$ was found to exist in Ca-GIC; unlike Li-and K-intercalation[8], to the best of our knowledge a lower stage $CaC_6$ compound has not been reported). Supplementary Figure S4 compares optical photographs of intercalated laminates after the exposure corresponding to saturation (such that no further colour changes occurred with further increase of exposure time) and after carrying out the intercalation process for approximately half the time (top and bottom rows, respectively). It is clear that the colours of GLs exposed to a metal for a shorter time are different from those intercalated to saturation, indicating lower carrier concentrations (see Supplementary Notes 8 and 9), presumably due to lower coverage of graphene crystallites with metal atoms. Accordingly, it should be possible, in principle, to continuously tune the level of doping and the associated electronic properties of graphene laminates, similar to an isolated MLG[11]. In practice however, at intermediate exposures it was difficult to achieve uniform colours (that is, uniform metal coverage and doping) over an entire GL sample; the colours shown in the bottom row of Fig. S4 were only found in some parts of a sample while other parts were either still dark grey (not intercalated) or of a yet another colour. Achieving an intermediate coverage was particularly difficult for Ca intercalation, as is clear from comparison of the corresponding images in Fig. S4.

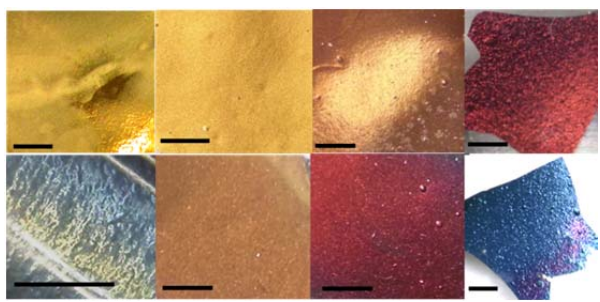

***Supplementary Figure S4.*** *Optical photographs of fully- and partially intercalated GLs (top and bottom rows, respectively). The metals used for intercalation, from left to right: Ca, Li, Cs and K. For example, the top left image is Ca-GL intercalated to saturation and bottom left image is Ca-GL after twice shorter intercalation time. All scale bars correspond to 1mm.*

Therefore we used an alternative approach and varied the carrier concentration by using mixed graphene-BN laminates (GBNLs), as described above, and exposing them to Ca metal until saturation was achieved. The colours of GBNLs with different BN content are shown in Fig. 3 of the main text. Corresponding Raman spectra (Supplementary Fig. S5) clearly indicate a decrease of the carrier density, $n$, in Ca-coated graphene crystallites within GBNLs as the proportion of BN in a laminate is increased (see also the reflectivity spectrum in Fig. 3e and the discussion below). Raman spectra taken on many different parts of each of these samples were identical, verifying sample homogeneity. Comparison of the three spectra in Supplementary Fig. S5 shows that the addition of 25% BN did not have a significant effect on $n$, with the same broad peak observed at ~1400 cm$^{-1}$ as for Ca-GL and Ca-RGOL (Supplementary Fig. S3). In contrast, the spectra for 50% and 70% BN show signatures of much less doped graphene (G peak at ≈1550-1600 cm$^{-1}$) and of hBN (1365 cm$^{-1}$ peak[24]). The lower level of Ca doping (presumably due to lower coverage with Ca atoms) is also evident from optical reflectivity measurements (main text and Supplementary Note 8).

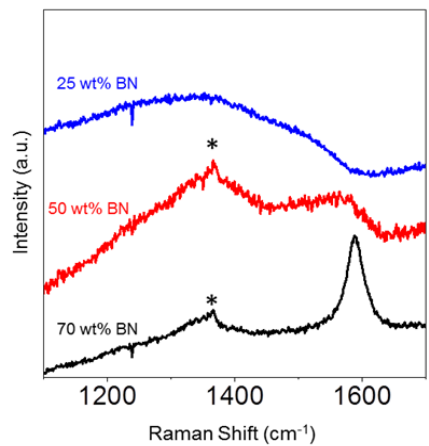

***Supplementary Figure S5.*** *Raman spectra of Ca-GBNLs with 25, 50 and 70 wt% BN content (\* indicates the peak corresponding to hBN).*

**Supplementary Note 8. Optical reflectivity.**

In addition to the reflectivity data presented in the main text, spectra were also taken for Li-, K- and Cs-intercalated GLs – see Supplementary Fig. S6a. To extract information about the electronic properties of the system (e.g., plasma frequency), we fit the experimental spectra with the well-known expression for the reflection coefficient, $R = \left|\frac{n-1}{n+1}\right|^2$, where the refractive index $n^2 \approx k\left(1 - \frac{\omega_p^2}{\omega(\omega - i/\tau)}\right)$ is derived from Maxwell's equations[25]. Here, $k$ is the dielectric permittivity of the environment, $\omega_p$ the plasma frequency, $\omega$ the frequency of the incident light and $\tau$ the electron collision time. The relatively shallow slopes of $R(E)$ curves in Figs. 3 and S6 indicate inhomogeneity of the electron distribution in the samples (for example due to slightly different coverage of individual graphene crystallites with metal atoms or different coverage at the edges of crystallites). To account for this, the fitting procedure allowed for a variation of $\omega_p$ within ±0.3-0.5 eV range – see caption to Fig. S6 for exact fitting parameters. As an example, Supplementary Figure S6b shows the experimental reflectivity spectrum for Ca-GL and the corresponding fitting curve; the extracted plasma frequency in this particular case was $\omega_p$ = 2.6 eV. Similar analysis was carried out for all other samples; the results are given in Supplementary Table 1. Notably, while Ca-GLs are superconducting and Li-GLs are not, both have the same plasmon energy, indicating similar overall electron concentrations. GLs intercalated with Cs and K also have plasmon energies similar to each other but significantly lower than for Ca- and Li-intercalated laminates. Compared to samples of intercalated bulk graphite that we prepared in parallel with GLs, all metal-doped GLs have significantly lower plasmon energies, again indicating lower overall electron concentrations in all laminate samples compared to intercalated graphite, presumably due to larger separations between graphene and metal atoms in the laminates compared to GICs.

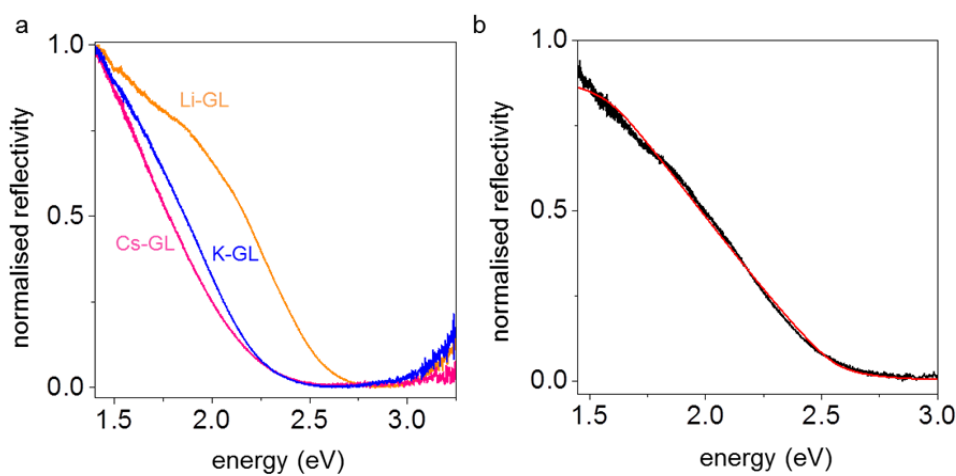

**Supplementary Figure S6**. *(a) Normalised reflectivity spectra of Li-, Cs- and K-intercalated GLs. (b) A normalised reflectivity spectrum of Ca-GL and the corresponding data fitting for the reflection coefficient, R(E).* The fitting parameters in this case were: $\Delta\omega_p$ =0.5 eV; $\tau$ =0.1; k =2.

| Sample | plasmon energy, E (eV) |
|---|---|
| Ca-GL | 2.6 |
| Ca-RGOL | 2.4 |
| Ca-GBNL (50 wt%) | 2.1 |
| Li-GL | 2.6 |
| Cs-GL | 2.3 |
| K-GL | 2.3 |
| Ca-GIC | 3.1 |
| Li-GIC | 3.0 |
| K-GIC | 2.5 |
| Cs-GIC | 2.5 |

**Supplementary Table 1.** *Plasmon energies of different intercalated GLs and corresponding bulk GICs estimated from the reflectivity spectra.*

## Supplementary Note 9. Calculations of charge carrier distribution in metal-intercalated graphene laminates

For a layered multicomponent system, such as metal-intercalated GLs, the knowledge of $\omega_p$ alone is not sufficient to extract quantitative information about the density of carriers and their distribution between graphene and the metal layers. To achieve this, we used a simple model where the metal-graphene layers are represented by a set of two two-dimensional units coupled electrostatically[26] and calculated the carrier densities in graphene and the metal layer (Inter layer, IL), such that they correspond to the experimentally found $\omega_p$. The electronic structure within each unit is given by three degenerate Dirac bands and one parabolic band located between the graphene and the metal layer[26]. A related situation, plasmons in a system consisting of a graphene layer and a two dimensional electron gas, has been considered in ref. [27]. Besides the electronic bands, other parameters included in the model are the width of the graphene layer, $d_C$, the width of the region occupied by the interlayer state, $d_{IL}$, the dielectric constant of the background, $\epsilon_0$, and the total number of carriers, $n_{total}$ - see Supplementary Fig. S7. The constraint of charge neutrality implies that the charge in the metal, $n_{total}$, satisfies $n_{total} = -n_C - n_{IL}$, where $n_C$ and $n_{IL}$ are the carrier densities in graphene and in the IL band, respectively.

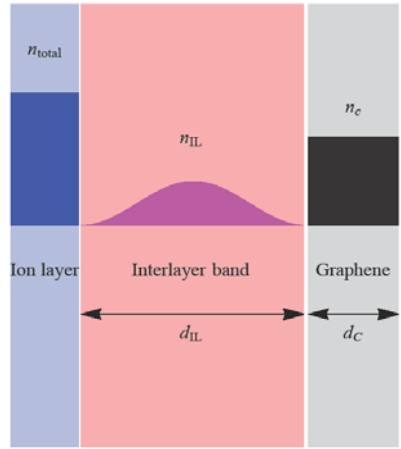

***Supplementary Figure S7.*** *Schematic representation of metal decorated graphene layers in GLs.*

The electronic bands (Supplementary Fig. S8) are

$$\varepsilon_D(\vec{k}) = E_D + v_F|\vec{k}|$$
$$\varepsilon_{IL}(\vec{k}) = E_{IL} + \frac{\hbar^2|\vec{k}|^2}{2m_{IL}}.$$

We take the energy at the Dirac point as $E_D = 0$. The minimum of the interlayer band is $E_{IL}$ and the Fermi energy is $E_F > E_{IL}, E_D$, so that the parabolic band is partially occupied, and the Dirac bands are *n* doped.

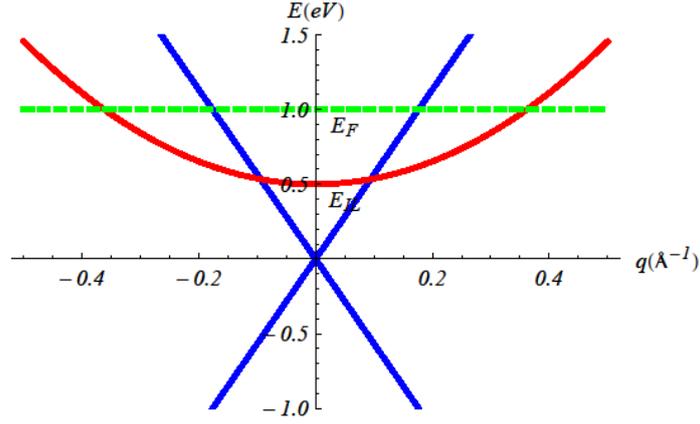

**Supplementary Figure S8.** Sketch of the electronic bands for metal decorated layers of graphene. Blue: three-fold degenerate graphene bands; red: interlayer band; green: Fermi level.

The properties of each metal-graphene compound in our study are determined by the total number of carriers, $n_{total}$. From $n_{total}$ the number of carriers in the graphene layer, $n_C$, and the IL band, $n_{IL}$, are derived by minimizing the total energy. Once the carrier distribution is determined, we calculate the plasmon frequency, $\omega_p$. The value of $n_{total}$ is then adjusted so that (i) the plasmon frequency agrees with our experimental reflectivity data and (ii) the value of $n_{IL}$ is zero in non-superconducting compound (Li-GL) and it is a monotonically increasing function of the superconducting critical temperature, $T_c$, in Ca-GLs, Ca-RGOL, Ca-GIC, as observed in our experiments. Due to the complexity of modelling a multilayer system containing both graphene and BN, we did not perform calculations for Ca-GBNLs.

As found experimentally, Ca-GL and Li-GL have the same plasmon energy, $\hbar\omega_p \cong$ 2.6 eV. We estimate $d_{IL}$ and $d_C$ from the experimentally determined interlayer separations, $d_{IL} + d_C$, in proportion to the known atomic radii of carbon and the dopant metal. For example, for Ca-GL $d_{IL} + d_C \approx 5.1$ Å, $d_{IL} \approx 3.7$ Å and $d_C \approx 1.4$ Å; for Li-GL $d_{IL} + d_C \approx 3.7$ Å, $d_{IL} \approx 2.5$ Å and $d_C \approx 1.2$ Å. We also consider Ca-GIC and Ca-RGOL where the interlayer distances are $d_{IL} + d_C \approx 4.5$ Å and 5.4 Å, respectively, and the corresponding plasmon energies, $\hbar\omega_p$ = 3.1 eV and 2.4 eV.

The distribution of carriers between the interlayer and graphene bands is determined by the electrostatic interactions, the quantum capacitance of each band and the position of $E_{IL}$ (Fig. S8). We assume that the charge in graphene, $n_C$, is distributed uniformly within a layer of thickness $d_C$,

$$\rho_C(x) = \begin{cases} 0 & x \leq -d_C \\ \dfrac{n_C}{d_C} & -d_C \leq x \leq 0 \\ 0 & 0 \leq x \end{cases}$$

The distribution of the charge in the IL band is

$$\rho_{IL}(x) = \begin{cases} 0 & x \leq 0 \\ \dfrac{n_{IL}}{2d_{IL}} \sin^2\left(\dfrac{\pi x}{d_C}\right) & 0 \leq x \leq d_{IL} \\ 0 & d_{IL} \leq x \end{cases}$$

The charge of the metal, $n_{total}$, is within a thin layer at $x = d_{IL}$. The electrostatic interactions per unit area between the metal ions and the carriers in graphene and the interlayer band are:

$$E_{el}^{ions-C} = e^2 n_C n_{total} \frac{d_C + 2d_{IL}}{16\pi}$$

$$E_{el}^{ions-IL} = e^2 n_{IL} n_{total} \frac{d_{IL}}{32\pi}$$

The electrostatic interactions between carriers are

$$E_{el}^{C-C} = e^2 n_C^2 \frac{d_C}{24\pi}$$

$$E_{el}^{IL-IL} = e^2 n_{IL}^2 \frac{d_{IL}(-3 + 2\pi^2)}{96\pi^2}$$

$$E_{el}^{C-IL} = e^2 n_C n_{IL} \frac{d_C + d_{IL}}{32\pi}$$

Finally, the quantum contribution of the carriers is:

$$E_q^C = \frac{2\hbar v_F}{\pi} \left(\frac{\pi n_C}{3}\right)^{3/2}$$

$$E_q^{IL} = E_{IL} n_{IL} + \frac{\hbar^2 \pi n_{IL}^2}{2m_{IL}}$$

The values of $n_C$ and $n_{IL}$ minimize the total energy

$$E_{tot} = E_{el}^{ions-C} + E_{el}^{ions-IL} + E_{el}^{IL-IL} + E_{el}^{C-C} + E_{el}^{C-IL} + E_q^C + E_q^{IL}$$

with a constraint $n_{total} = -n_C - n_{IL}$. For $E_{IL} > 0$ and small values of $n_{total}$ all carriers reside in the graphene layer. As $n_{total}$ increases, carriers move into the IL band.

The total carrier concentrations were optimised as shown by labels in Supplementary Fig. S9, e.g., $\cong 1.8 \times 10^{14}$ cm$^{-2}$ for, Ca-GIC, $1.1 \times 10^{14}$ cm$^{-2}$ for Ca-GL and so on. Fig. S9 shows the total energy for different modelled structures as a function of $n_C$, at fixed $n_{total}$. We assume that the position of the interlayer band does not vary significantly among the different compounds. We choose $E_D = 0$, and $E_{IL} = 0.5$ eV, so that, at low total carrier density the carriers are in the graphene layer, and move into the interlayer band as the concentration increases.

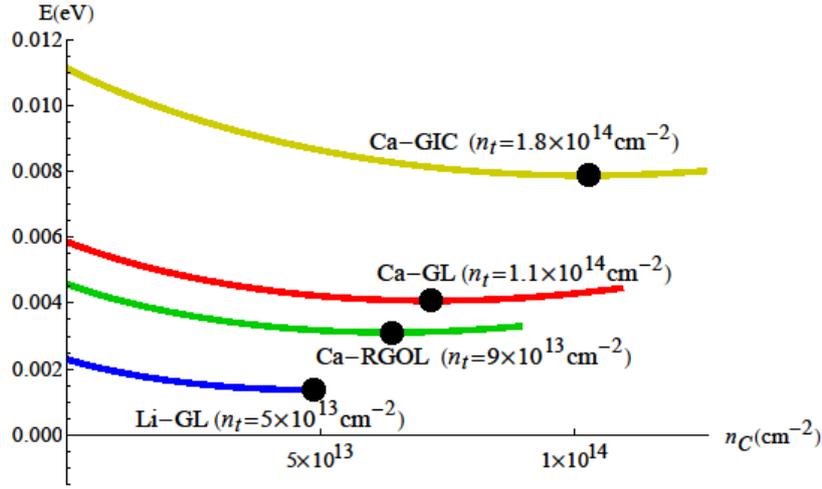

***Supplementary Figure S9.*** *Total energy of the modelled structures as a function of the carrier density in graphene layers. Black dots indicate the minimum total energy and the labels give corresponding values of $n_{total}$ with ~20% accuracy. The Fermi velocity of the Dirac bands is that of graphene, and the effective mass of the parabolic band is $m_{IL} = m_e$. The remaining parameters are given in the text.*

Minimization of $E_{tot}$ allowed us to find the distribution of carriers between graphene and the IL bands. The obtained contributions of $n_{total}$ to the graphene bands are ≈ 1.0×10$^{14}$, 7.2×10$^{13}$ and 6.4×10$^{13}$ cm$^{-2}$ for the modelled Ca-GIC, Ca-GL and Ca-RGOL, respectively, i.e. bulk Ca-GIC has approximately 30% higher electron concentration in graphene layers compared to Ca decorated graphene crystallites in GLs.

The plasmon energy for one graphene-IL band unit can be calculated from the polarization, $\chi(\vec{q}, \omega)$. At small wave vectors $\vec{q}$ it is

$$\chi(\vec{q}, \omega) = 3 \times \chi_C(\vec{q}, \omega) + \chi_{IL}(\vec{q}, \omega) \approx \frac{3 v_F k_F^C |\vec{q}|^2}{\pi \hbar \omega^2} + \frac{2(k_F^{IL})^2 |\vec{q}|^2}{2\pi m_{IL} \omega^2}$$

The plasmon frequency, $\omega_p$, is given by the solution of the equation

$$1 = v_{\vec{q}}\, \chi(\vec{q}, \omega_p^2)$$

where $v_{\vec{q}} = (2\pi e^2)/(\epsilon_0 |\vec{q}|)$ is the Coulomb potential, and

$$n_{IL} = \frac{(k_F^{IL})^2}{2\pi}$$

$$n_C = \frac{3(k_F^C)^2}{\pi}$$

In terms of the carrier densities $n_C$ and $n_{IL}$, the plasmon frequency is

$$\hbar^2 \omega_p^2(\vec{q}) = \frac{e^2 |\vec{q}|}{\epsilon_0} \left( \frac{2\pi \hbar^2 n_{IL}}{m_{IL}} + 2\sqrt{3\pi} \hbar v_F \sqrt{|n_C|} \right)$$

This calculation can then be extended to a periodic stack of the two-dimensional graphene-IL units. In this case the plasmon frequency obeys

$$1 = v_{\vec{q}}\, \chi(\vec{q}, \omega_p^2) \frac{\sinh(|\vec{q}|d)}{\cosh(|\vec{q}|d) - \cos(k_z d)}$$

where $d$ is the spacing of the units and $k_z$ is the wavevector along the direction normal to the units. For $|\vec{q}|d, k_z d \to 0$ the plasmon frequency is

$$\hbar^2 \omega_p^2 = \frac{2e^2}{\epsilon_0 d}\left(\frac{2\pi \hbar^2 n_{IL}}{m_{IL}} + 2\sqrt{3\pi}\hbar v_F \sqrt{|n_C|}\right)$$

This expression shows that the plasmon frequency depends not only on the total carrier density (as would be the case for a 3D metal), but also on how the carriers are distributed between the graphene layer and the IL band. Once the carrier distribution is determined by minimizing the energy (Fig. S9), the plasmon frequencies can be calculated using the above expression. Supplementary Fig. S10 shows the dependence of plasmon energy, $\hbar\omega_p$, on the carrier concentration in the IL band, $n_{IL}$.

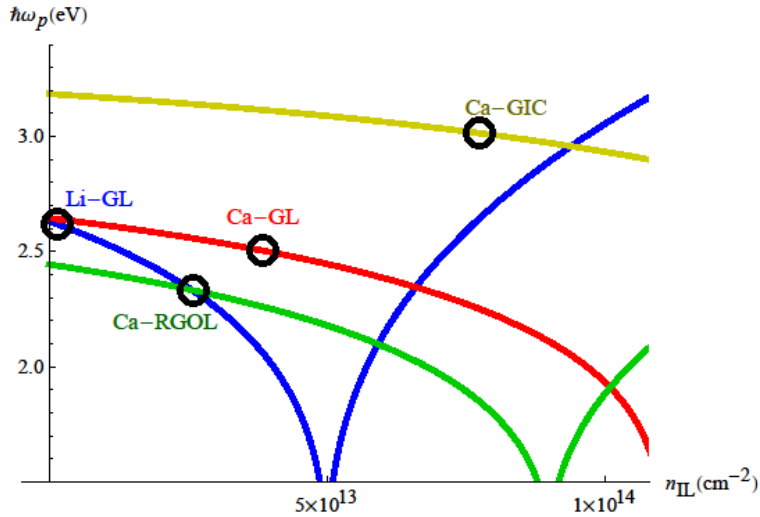

***Supplementary Figure S10**. Plasmon energy of an infinite stack of layers with one parabolic band and three Dirac bands as a function of the carrier density in the interlayer band, $n_{IL}$. The round symbols indicate the value of $n_{IL}$ corresponding to the experimentally determined plasmon energy.*

Analysis of Fig. S10 clearly shows that it is possible for different metal-intercalated compounds to have the same plasmon energy (as is the case for Li-GL and Ca-GL) but different distributions of charge carriers between the IL and graphene bands, and as a result different superconducting properties. Furthermore, for superconducting metal-intercalated graphene compounds, their $T_c$ depends on the level of doping of the graphene layers, which, in its turn, is determined by the total density of charge carriers in the system. That is, compounds with lowest total carrier densities (in our case, RGOLs and GBNLs with 70% BN) are also expected to have lowest carrier concentrations in graphene layers and, accordingly, lowest $T_c$'s, in agreement with the observations.

Finally, we can use the estimated differences in carrier concentrations in different superconducting samples (Ca-GIC, Ca-GLs and Ca-RGOLs) to estimate the changes in the critical temperature, $T_c$, that would be expected if this were the only factor affecting $T_c$. As was shown in ref. [28], variations in the density of states, $\delta N(0)$ lead to variations in $T_c$ according to

$$\frac{\delta T_c}{T_{c0}} = \frac{1}{N(0)V}\frac{\delta N(0)}{N(0)},$$

where

$$N(0)V = -1/\ln\left(\frac{T_{c0}}{1.14\theta_D}\right).$$

Here $T_{c0}$ is the initial $T_c$, $\theta_D$ is the Debye temperature and $V$ the electron-phonon coupling strength. As the density of states of the interlayer band does not depend on the filling, we can assume that $\delta N(0)$ is proportional to variations of the carrier concentration in doped graphene layers, i.e.

$$\frac{\delta N(0)}{N(0)} \approx \frac{\Delta n_c}{n_c}$$

or

$$\Delta T_c(\Delta n) \approx T_{c0}\frac{\Delta n_c}{n_c}\ln\left(\frac{T_{c0}}{1.14\theta_D}\right)$$

Using $\theta_D$ = 175K[17] and $\Delta n_c/n_c \approx 10\%$ calculated as shown above from the measured plasmon energies for Ca-GL and Ca-RGOL, we expect $T_c^{Ca-RGOL}$ to be 2.2K lower than $T_c^{Ca-GL}$, in excellent agreement with the observed difference of 2.4K. However, a similar comparison of Ca-intercalated bulk graphite ($n_C \approx 1.0 \times 10^{14}$ cm$^{-2}$) and Ca-GL ($n_C \approx 7.2 \times 10^{13}$ cm$^{-2}$) gives a large overestimate of the expected decrease in $T_c$, indicating that other factors must be taken into account as explained in the main text.

**Supplementary references**